\documentclass{pasj00}

\SetRunningHead{Morihana et al.}{MAXI/GSC Discovery of MAXI~J1305$-$704}

\Received{2013 July 8}

\usepackage{natbib}
\usepackage{times}
\usepackage[dvips]{graphicx}

\newcommand{\ergs}{{\rm erg}\ {\rm s}^{-1}}
\newcommand{\ergcms}{{\rm erg}\ {\rm cm}^{-2}\ {\rm s}^{-1}}

\begin{document}

\title{MAXI/GSC Discovery of the Black Hole Candidate MAXI~J1305$-$704}
\author{
  Kumiko \textsc{Morihana}, \altaffilmark{1,2}
  Mutsumi \textsc{Sugizaki}, \altaffilmark{2}
  Satoshi \textsc{Nakahira}, \altaffilmark{3}
  Megumi   \textsc{Shidatsu}, \altaffilmark{4}
  Yoshihiro \textsc{Ueda}, \altaffilmark{4}
  Motoko \textsc{Serino}, \altaffilmark{2}
  Tatehiro \textsc{Mihara}, \altaffilmark{2}
  Masaru  \textsc{Matsuoka}, \altaffilmark{2}
   Hitoshi \textsc{Negoro}, \altaffilmark{5}
  Nobuyuki \textsc{Kawai}, \altaffilmark{6}
  and the MAXI team
}
\altaffiltext{1}{Nishiharima Astronomical Observatory, Center for Astronomy, University
of Hyogo, 407-2 Nichigaichi, Sayo-cho, Sayo, Hyogo, 670-5313}
\altaffiltext{2}{MAXI team, Institute of Physical and Chemical Research (RIKEN), 2-1 Hirosawa, Wako, Saitama 351-0198}
\altaffiltext{3}{ISS Science Project Office, Institute of Space and Astronautical
Science, Japan Aerospace Exploration Agency, 2-1-1 Sengen, Tsukuba,
Ibaraki 305-8505}
\altaffiltext{4}{Department of Astronomy, Kyoto University, Oiwake-cho, Sakyo-ku, Kyoto
606-8502}
\altaffiltext{5}{Department of Physics, Nihon University, 1-8-14 Kanda-Surugadai, Chiyoda-ku, Tokyo 101-8308}
\altaffiltext{6}{Department of Physics, Tokyo Institute of Technology, 2-12-1 Ookayama,
Meguro-ku, Tokyo 152-8551}

\email{morihana@nhao.jp}

\maketitle

\begin{abstract}
We present the first results on the new black hole candidate, MAXI
J1305$-$704, observed by MAXI/GSC.  The new X-ray transient, named as
MAXI J1305$-$704, was first detected by the MAXI-GSC all-sky survey on
2012 April 9 in the direction to the outer Galactic bulge at
$(l,b)=$(\timeform{304D.2},\timeform{-7D.6}). The Swift/XRT
follow-up observation confirmed the uncatalogued point source and
localized to the position at (\timeform{13h06m56s.44},
\timeform{-70D27'4''.91}). The source continued the
activity for about five months until 2012 August. 
The MAXI/GSC light curve in the 2--10 keV
band and the variation of the hardness ratio of the 4--10 keV to the
2--4 keV flux revealed the hard-to-soft state transition on the the
sixth day (April 15) in the brightening phase and the soft-to-hard
transition on the $\sim$60th day (June 15) in the decay phase.  The
luminosity at the initial hard-to-soft transition was significantly
higher than that at the soft-to-hard transition in the decay phase.
The X-ray spectra in the hard state are represented by a single
power-law model with a photon index of $\sim 2.0$, while those in the
soft state need such an additional soft component as represented by a
multi-color disk blackbody emission with an inner disk temperature
$\sim$0.5--1.2 keV.  All the obtained features support the source
identification of a Galactic black-hole binary located in the Galactic
bulge.

\end{abstract}

\medskip

\section{Introduction}\label{s1}
Galactic Black Hole candidates in Binary systems (BHBs) occasionally
exhibit an explosive X-ray brightening referred to as ``X-ray nova''
or ``outburst'', which is triggered by a sudden increase of in-flowing
matter.  
A ``typical'' outburst light curve consists of a fast rise (a few
days) followed by a longer (a few months) exponential decline 
(see \citealt{Tanaka96} for review). 
However, recent RXTE and MAXI surveys
discovered a variety of outbursts with peculiar light-curve profiles
(e.g. \citealt{McC06,Nakahira10,Yamaoka12}).

During the BHB outbursts, the luminosity can vary by a factor of
~10$^{4}$--$10^{5}$ and also the spectrum will change
(see \citealt{McC06} for review).
These outbursts normally begin with the hard state. The hard-state spectrum is
approximated by a single power-law with a photon index,
$\Gamma=$1.4--2.0 up to the high-energy cutoff at $\sim 100$ keV and
interpreted as a thermal Comptonization of soft photons by hot and
optically thin electron cloud (e.g. \citealt{Takahashi08};
\citealt{Makishima08}).
During the initial brightening phase, they will exhibit the
hard-to-soft state transition.
The soft-state spectrum is represented by a combination of a
thermal blackbody emission from an optically-thick and geometrically-thin
accretion disk, so-called ``standard disk'' \citep{Shakura73},
with an inner-disk temperature, $T_{\rm in}\sim 1$ keV, and a
power-law with $\Gamma\sim$2--2.5.
In the course of the outburst decay phase, the source will return to
the hard state again.  
%
The hysteresis in the X-ray intensity during the state transitions
\citep{Miyamoto95} produces a q-shaped track in a hardness-intensity
diagram (HID), so called ``q-curve'' \citep{Homan05}.  The ``q-curve''
can be used to identify a new X-ray source as a BHB.

The new X-ray transient, MAXI~J1305$-$704, was first detected by MAXI
(Monitor of All-sky X-ray Image; \citealt{Matsuoka09}) GSC (Gas Slit
Camera; \citealt{Mihara11,Sugizaki11}) in the direction to the outer
Galactic bulge at $(l,b)=$(\timeform{304D.2},\timeform{-7D.6}) or (RA,
Dec)$=$(\timeform{196D.4}, \timeform{-70D.4}).  The source emergence
was first noticed by MAXI transient alert system \citep{Negoro12} at
2012 April 9 11:24:23 UT 
when the flux reached
$30\pm 18$~mCrab in the 4--10 keV band \citep{Sato12}.
%
%
%
Swift follow-up observations which covered the entire MAXI/GSC error
circle with the five pointing observations, confirmed an uncatalogued
bright X-ray source by both the X-Ray Telescope
(XRT:~\citealt{Gehrels05}) and the UVOT \citep{Roming05}, and then
localized the position at (RA,Dec)$=$(\timeform{13h06m56s.44},
\timeform{-70D27'04''.91}) with an uncertainty of 5\arcsec
~\citep{Greiner12, Kennea12a}.  
Optical follow-up observation found a new point source 
within the XRT error circle \citep{Greiner12}. 
From the X-ray and optical spectra as well as their time variation, 
the source was proposed to be a BHB
\citep{Greiner12,Kennea12a, Suwa12}.  Observed X-ray dip or
eclipse-like features \citep{Kennea12c,Kennea12b} and possible
absorption-line signatures \citep{Miller12a,Miller12b} suggest that
our line of sight locates relatively edge-on of the accretion disk.

In this paper, we present results of MAXI/GSC observations of
MAXI~J1305$-$704 covering the entire outburst period from 2012 April
to 2012 August, and discuss the identification and the nature based on
the light curve and the spectral variation.
In the following sections, all the quoted errors are given at the
90\% confidence limit.

\begin{figure}
 \begin{center}
 \FigureFile(80mm,){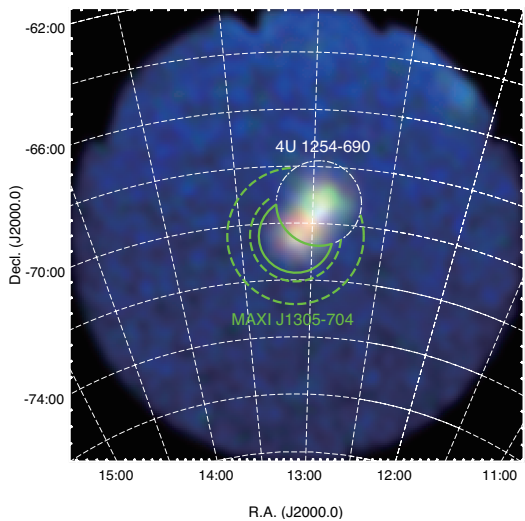}
  \end{center}
 \caption{ 
   MAXI/GSC false-color image integrating data taken from 2012
   April 3 to July 17 in 2--16~keV (Red: 2--4~keV, Green: 4--8 ~keV,
   and Blue: 8--16~keV). 
   A while solid circle is the excluded region to avoid contamination
   of 4U~1254$-$690, and a green circle is a typical
   source-extract region with a radius of $1.4^\circ$.  An annulus
   between green dashed lines is for the background region.}\label{f1}
\end{figure}

\section{Observation \& Results}\label{s2}
\subsection{X-ray Light Curves and Hardness Ratios}\label{s2-1}

We performed the GSC data analysis following the standard analysis
procedure described in \citet{Sugizaki11}.  Figure~\ref{f1} shows the
GSC image accumulating data taken from 2012 April 3 to July 17 in a
region of \timeform{5D} from MAXI J1305$-$704.  As seen in the image, a
nearby bright X-ray source, 4U 1254$-$690, is located at a distance of
\timeform{1D.4}. To avoid the contamination from the source,
we carefully selected the source and the background events.
We first excluded data within
$\sim$\timeform{1D.6} from 4U 1254$-$690, then collected source events from
a circle of $\sim$\timeform{1D.4} centered at the target position. To
optimize the source-to-background event ratio, we adjusted these areas
in each scan transit.
Background events were then collected from the source free region
within \timeform{2D.5} from the MAXI J1305$-$704 position.  

\begin{figure}
 \begin{center}
   \FigureFile(80mm,){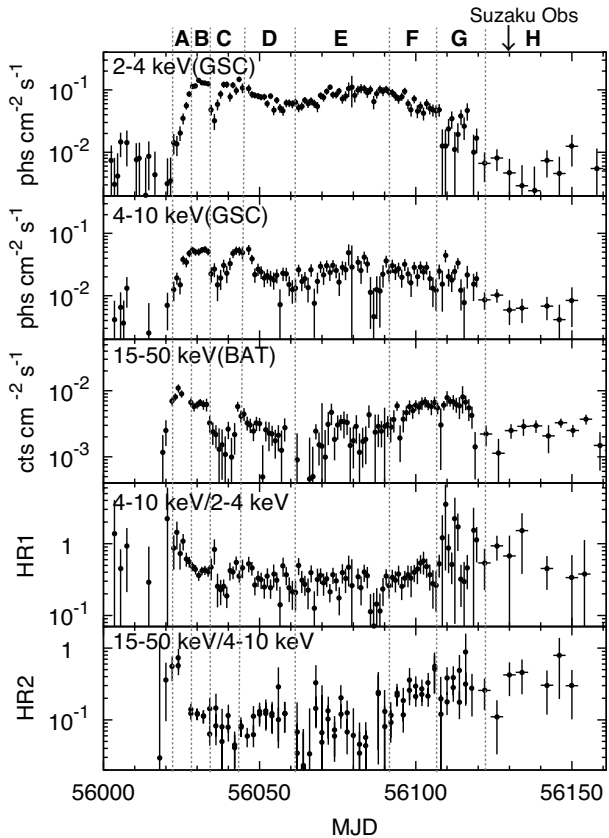}
 \end{center}
 \caption{ MAXI/GSC and Swift/BAT light curves of MAXI~J1305$-$704
   from MJD$=$56000 (2012 March 14) to MJD$=$56160 (2012 August 21) in
   three energy ranges of 2--4~keV, 4--10~keV (GSC), and 15--50~keV (BAT) , and
   their hardness ratios. The data are binned per a day in
   MJD$=$56000--56120, and per four days in MJD$=$56120--56160.  
   Vertical error bars represent the 1$\sigma$ statistic uncertainty.  The
   epoch of Suzaku observation \citep{Shidatsu13} is also indicated.
 }\label{f2}
\end{figure}

To obtain light curves, we calculated a photon flux corrected for
effective-area variation in each scan transit in each energy band.  In
the GSC data, we did not find any significant flux in a light curve
above 10 keV in time bins from 1 d to 4 d. Figure~\ref{f2} shows the
obtained light curves in 2--4 keV and 4--10 keV, and that of the
Swift/BAT in 15--50 keV obtained from the Swift/BAT transient-monitor
archive\footnote{http://heasarc.gsfc.nasa.gov/docs/swift/results/transients/}
provided by the Swift/BAT team.  To clarify the spectral variation, we
also plot two hardness ratios of the 4--10 keV flux to the 2--4 keV
flux (HR1) and the 15--50 keV to the 4--10 keV (HR2) in figure
\ref{f2}.  We divided the entire outburst period into eight intervals,
A, B, C, ..., and H, as illustrated in figure \ref{f2}, to investigate
detail flux and spectral variations in each outburst phase.

The light curve apparently shows different profiles among the three
energy bands.  
In the initial brightening phase (period A), the onset is apparently
earlier in the higher energy band.  Both the two hardness ratios, HR1
and HR2, then decreased during the period A.  This indicates that the
hard-to-soft state transition occurred in the initial phase
\citep{Suwa12}.
%
%
After the 2--4 keV flux peaked at MJD$\simeq$56030 (period B), the source 
had stayed
an active phase with a flux above 30 mCrab for $\sim 60$ days until
MJD$\simeq$56110 (period G), where the HR1 had kept at $\simeq 0.3$.
The source activity then began to decrease gradually since
MJD$\simeq$56120 (period G). During the outburst decay phase, the flux
decline started earlier in the lower energy band, and then the two
hardness ratios increased. This implies that the soft-to-hard state
transition would occur.

The spectral changes are most apparent in the HR1.  We thus utilize
the HR1 value to discriminate the soft and the hard states, hereafter,
and set their boundary at HR1$=0.5$. This classifies the periods A and
G into the hard state, and the others into the soft state.

\subsection{Hardness-Intensity Diagram (HID)}\label{s2-2}
\begin{figure}
 \begin{center}
  \FigureFile(80mm,){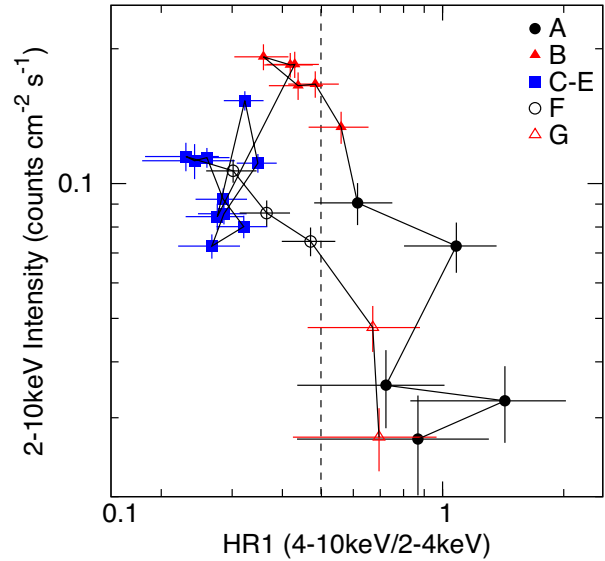}
 \end{center}
 \caption{ 
   Hardness-intensity diagram during the outburst from
   MJD$=$56022 to MJD$=$56120. Data in the periods A, B, C--E, F, and
   G are marked with black filled circle, red open triangle, blue filled
   box, black open circle and red open triangle, respectively.  Each
   data point represent the 1-day average in the periods A and B, and
   the 5-day average in the periods C--E, F, G.  Error bars represent
   the 1-$\sigma$ statistical uncertainty.  }\label{f3}

\end{figure}

In figure~\ref{f3}, we plot a HID representing the relation between
the 2--10 keV flux and HR1 during the outburst from MJD$=$56022 to
MJD$=$56120.  According to the photon count statistics of each
outburst phase, data are binned by one-day intervals in the periods A
and B, and by 5-day intervals in the periods C, D, E, F, and G.
Due to the luminosity hysteresis in the state transitions in both the
brightening and the decay phases, the HID exhibits a q-shaped track,
which is a common feature in BHBs \citep{Homan05}.

\subsection{Spectral Analysis}\label{s2-3}

The q-shaped HID track obtained in the previous subsection suggests
that MAXI J1305$-$704 is possibly a BHB.  We thus performed spectral
analysis by fitting data with typical BHB emission models.
We discuss validity of the assumed models and the source nature from
the obtained best-fit parameters.

\begin{figure*}
 \begin{center}
 \FigureFile(80mm,){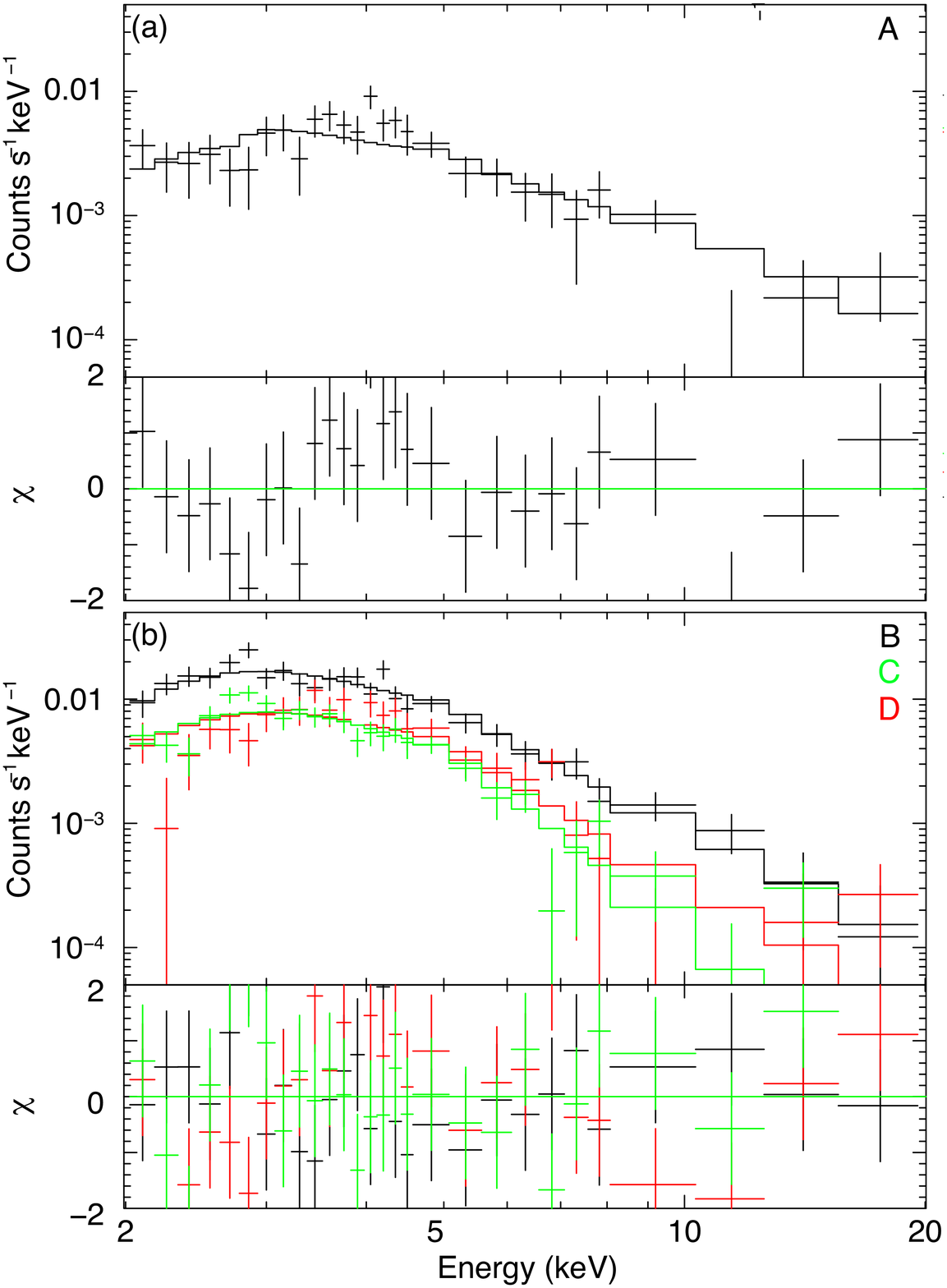}
 \FigureFile(78mm,){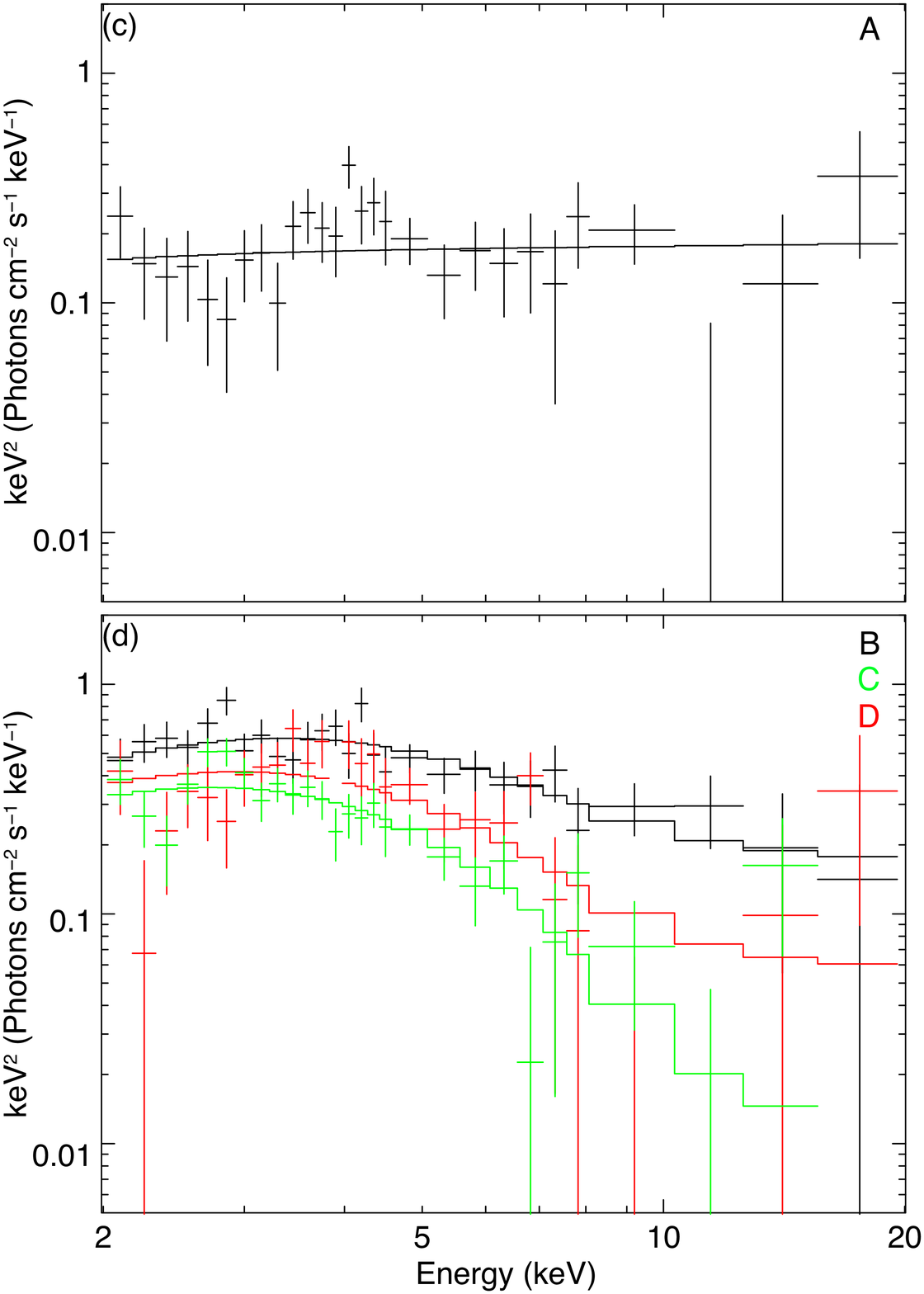}
 \end{center}
 \caption{Examples of MAXI J1305$-$704 spectra by MAXI/GSC in (a)
   hard-state period A, and in (b) soft-state periods B, C, and D. The
   best-fit {\tt phabs*powerlaw} model for the hard-state (a) and the
   {\tt phabs*simpl*diskbb} models for the soft-state (b), folded with
   the instrument response, are overlaid. The data-to-model residuals
   ($\chi$) are also shown in each panel.  Unfolded $\nu f\nu$ spectra
   of these hard and soft-state periods are presented in panels (c)
   and (d), respectively.  }\label{f4}
\end{figure*}

GSC spectral data and their response files were obtained by using the
MAXI on-demand data system\footnote{http://maxi.riken.jp/mxondem/}
(\citealt{Nakahira12a}). The system implemented event data version 1.4
containing data taken by the GSC counters operated at both the 1550 V
and the 1650 V. The total exposure reached 220.6 ks cm$^2$ with 1172
transits.  Spectral data were binned so that each bin contains at
least $\sim 30$ photons from the source region.  All spectral fits
were carried out on XSPEC version 12.7.1.  Figure~\ref{f4} shows the
obtained spectra in the periods A, B, C, and D as examples.

We employed a single power-law model for the hard-state emission
spectra, and a partly Comptonized multi-color-disk (MCD) model
\citep{Mitsuda84, Makishima86}, represented by {\tt simpl*diskbb} in
Xspec terminology, for the soft-state spectra.  The \texttt{simpl} is
an empirical Comptonization model which converts a given fraction
$f_{\rm sc}$ of incident photons into a power-law shape with a given
photon index $\Gamma$ (\citealt{Steiner09}).  We fixed the $\Gamma$ of
the {\tt simpl} model at a typical BHB soft-state value, 2.2
\citep{Ebisawa94, Kolehmainen11}, and accounted only the up-scattered
component.  Since the MCD dominates the GSC energy band, the data does
not have enough statistics to determine the $\Gamma$ in the soft state.
To take account of the interstellar absorption, {\tt phabs} model with an
abundance of \cite{Anders89} was applied.
We fixed the hydrogen column density, $N_{\rm H}$, at the Galactic H$_{\rm I}$
density, $0.18\times 10^{22}$ cm$^{-2}$, in the source direction
\citep{2005A&A...440..775K}.
%
Such a column density as the Galactic H$_{\rm I}$ affects the GSC energy band little.

Table~\ref{t01} summarizes the best-fit parameters obtained for the
the periods A to G.  All the fits were accepted within the 90\%
confidence limits.  In figure~\ref{f4}, 
the folded/unfolded best-fit models and their
data-to-model residuals are shown on each observed spectrum.
In the hard state, the power-law photon index are always
$\sim$2.0, which agrees with a typical value in the BHB hard states.
In the BHB soft state, the inner-disk radius is supposed to become
constant when it reach the innermost stable circular orbit.  As seen
in Table~\ref{t01}, the MCD normalization parameter $r_{\rm in}
\sqrt{\cos i}$ is almost consistent with each other in the three
soft-state periods B, C, and D, where the inner disk temperature
$T_{\rm in}$ is also stable at $\sim$1~keV.  We thus fitted these
soft-state spectra simultaneously, with a common inner-disk
radius. The fit was acceptable and the the obtained best-fit
parameters are shown together in Table~\ref{t01}.
The $r_{\rm in}\sqrt{\cos i}$ value then increased in the period E and
F. This is considered to reflect the fact that the soft-to-hard state
transition started gradually, as has been seen in the HID
(figure~\ref{f3}).

\begin{table*}
\scriptsize
  \caption{Best-fit model paramters obtaibed from MAXI/GSC 2--20~keV spectra.}\label{t01} 
  \begin{center}         
  \begin{tabular}{cccccccclll}
  \hline\hline
    Period  & MJD & Exposure &$T_\mathrm{in}$
   &$r_\mathrm{in}$~$\sqrt{\cos i}$\footnotemark[$*$]  
& $\Gamma$ &  $f_\mathrm{sc}$\footnotemark[$\dagger$] &  $F_\mathrm{t}$\footnotemark[$\ddagger$]&
   $F_\mathrm{pl}$\footnotemark[$\ddagger$]  & $F_\mathrm{db}$\footnotemark[$\ddagger$] & $\chi^{2}$~/~d.o.f\\
     & & (ks$\cdot$cm$^{2}$)&　(keV) &(km) & & & & &  &\\
   \hline
   A & 56022.0--56028.0 & 20.0 &--& --& $2.0_{-0.2}^{+0.3}$  & -- & $6.3_{-0.6}^{+0.3}$
			       & 6.3 & -- &
   30.0 / 26\\
   B & 56028.0--56034.0 & 13.8 &  $0.98_{-0.24}^{+0.25}$&  $9.6_{-3.4}^{+7.0}$  & 2.2
   (fix) & $0.30_{-0.13}^{+0.14}$ & $15.1_{-0.5}^{+0.2}$  & 9.4  & 5.6 &  17.9 / 28\\
   C & 56034.0-56042.0 & 17.7 & $1.41_{-0.17}^{+0.18}$&$3.8_{-2.5}^{+3.2}$  &  2.2 (fix) &
   $0.00_{-0.0}^{+0.09}$ & $7.9_{-0.7}^{+0.5}$ & 0.0 & 7.9 &  33.7 / 25\\ 
   D & 56042.0--56060.0 & 48.1 & $0.96_{-0.14}^{+0.16}$ &  $8.2_{-5.7}^{+8.3}$  & 2.2 (fix) &
   $0.08_{-0.08}^{+0.09}$ & $6.8_{-1.0}^{+0.8}$ & 1.8 & 4.9  &  31.4 / 25 \\ 
   E & 56060.0--56090.0 & 62.5 & $0.65_{-0.09}^{+0.10}$ &  $14.5_{-3.8}^{+5.7}$ & 2.2
		       (fix) &  $0.10_{-0.04}^{+0.04}$ & $7.4_{-0.2}^{+0.4}$ & 3.3 & 4.1 & 23.2 / 21 \\ 
   F & 56090.0--56105.0 & 34.9 & $0.46_{-0.16}^{+0.25}$ & $37.5_{-13.2}^{+62.5}$ 
   & 2.2 (fix)  &$0.18_{-0.10}^{+0.11}$  &  $8.2_{-3.1}^{+1.3}$ & 6.8  & 1.6 &  20.9 / 21\\
　 G & 56105.0--56120.0 & 23.6 & -- & --& $2.0_{-0.4}^{+0.4}$ & -- & $5.4_{-0.3}^{+0.8}$  &  5.4 & -- 
   & 11.4~/~10\\
   \hline
   B\footnotemark[$\S$]  & --  & --  & $1.21_{-0.14}^{+0.13}$ & $6.2_{-3.6}^{+5.0}$ &
		       2.2 (fix)
      & $0.21_{-0.10}^{+0.11}$ & $14.2_{-1.6}^{+1.0}$ & 6.4 &  7.8 &  88.8 / 75   \\
   C\footnotemark[$\S$] & --  &--   & $1.13_{-0.13}^{+0.13}$ & $=$(B)  & 2.2
   (fix) & $0.09_{-0.09}^{+0.14}$ &  $8.4_{-0.5}^{+0.6}$  & 2.1 & 6.2 &  $=$(B) \\
   D\footnotemark[$\S$] & --  & --  & $1.09_{-0.11}^{+0.10}$ & $=$(B) & 2.2 (fix)
		       &$0.02_{-0.02}^{+0.08}$ & $6.1_{-0.6}^{+0.1}$  & 0.4  & 5.7 &  $=$(B)  \\
   \hline
  \end{tabular}
  \end{center}
 \footnotemark[$*$]{Square root of the {\tt diskbb} model normalization, which is represented with  the inner disk radius $r_{\rm in}$ (km), the disk inclination $i$ and the source distnce $d_{10}$ (10 kpc) as $r_{\rm in}\sqrt{\cos i} / d_{10}$.  Values are given for an assumed source distance of 10 kpc.}\\
    \footnotemark[$\dagger$]{Scattering fraction of {\tt simpl} model}\\
    \footnotemark[$\ddagger$]{Absorption corrected model flux ($F_\mathrm{t}$) in units of
 $10^{-10}$~$\ergcms$ in 2--20~keV band, and fluxes of individual {\tt powerlaw} ($F_\mathrm{pl}$) and {\tt diskbb} ($F_\mathrm{db}$) components 
at the best fit.}\\
    \footnotemark[$\S$]{Spectra of periods B, C, and D are fitted simultaneously.}
\end{table*}

\section{Discussion}\label{s3}

We have analyzed the light curve and the spectral variation of the new
X-ray transient, MAXI J1305$-$704, obtained from the
MAXI/GSC all-sky survey data in the 2--20 keV band.
We here consider the identification and the nature of the new object 
from all the obtained results.

The MAXI/GSC light curves and the hardness variations in figure
\ref{f2} reveal the hard-to-soft state transition in the brightening
phase and the soft-to-hard transition in the decay phase.  The
q-shaped HID track in figure \ref{f3} clarifies the luminosity
hysteresis in the state transitions.  These observed behaviors agree
with the typical BHB feature \citep{Homan05}.
\cite{Gier06} suggested that the initial hard-to-soft transitions of
BHBs are divided into two types; ``bright/slow'' type which occurs at
the 30 \% of Eddington luminosity and takes more than 30 days, and
``dark/fast'' type which occurs at less than 10 \% of Eddington
luminosity and takes less than 15 days.  In the latter ``dark/fast''
type, the intermediate/very high state during the transition passes
quickly.  Observed transition on MAXI J1305$-$704, which had proceeded
for $\lesssim10$ days in the periods A$+$B, agree with the
``dark/fast'' type.
The 2-20 keV X-ray spectra, represented by a simple power-law with $\Gamma=$2. 
in the hard state and by a partly Comptonized
MCD model with the $T_{\rm in}=$0.5-1.2 keV and the scattering fraction
$f_{\rm sc} \simeq 0.2 $ in the soft state, also agree with the
typical BHBs \citep{McC06}. 
However, the similar behaviors have been observed in some of transient
neutron-star (NS) low-mass X-ray binaries
(e.g. \citealt{2006csxs.book...39V}).
Because of the limited photon statistics of the GSC data, a little
difference in X-ray spectra between the BH and the NS systems
cannot be distinguished.
Also, any confident evidence as a NS binary, such as a coherent
pulsation and type-I X-ray bursts, has not been detected.
Therefore, we cannot conclude if the object is either a 
BHB or a NS binary from the MAXI/GSC data alone.

\begin{figure}
 \begin{center}
 \FigureFile(75mm,){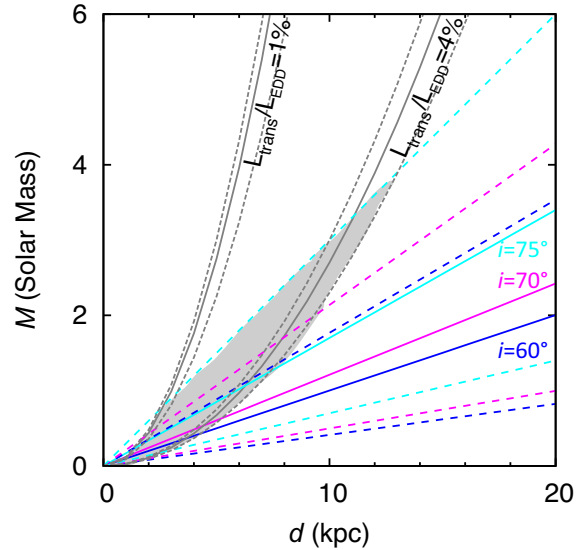}
 \end{center}
 \caption{Constraints on the mass-distance relation of MAXI
   J1305$-$704 from MAXI/GSC data.  Relations expected from the
   assumptions of $R_{\rm in}=R_{\rm ISCO}$ in the soft state and
   $i=60^\circ, 70^\circ, 75^\circ$, and of $L_{\rm trans}/L_{\rm EDD}
   = 1,4$\% are drawn with solid lines, and their limits within the
   statistical errors are in dashed lines.  A possible mass-distance
   region from the assumptions of $i\lesssim 75^\circ$ and 
   $1\%<L_{\rm trans}/L_{\rm EDD}<4\%$ is shadowed.}
 \label{f5}
\end{figure}

If MAXI~J1305$-$704 is a BHB, 
the mass of the central accretor should be higher 
than the maximum NS mass, at least $2\MO$.
We consider the constraint on the mass $M$ and the source distance $d$
relation by following the discussion in \citet{Yamaoka12} and \citet{Nakahira12}.
In the soft state, the inner radius of the accretion disk is supposed
to reach that of the innermost stable circular orbit, $R_{\rm ISCO}$,
which is the three times Schwaltzschild radius $=6GM/c^{2}$ if
the central accretor has no spin and $<6GM/c^{2}$ if it has a spin.
%
%
By using the MCD model parameter, the realistic estimate of the
innermost radius $R_\mathrm{in}$ is represented by
\begin{equation}
R_\mathrm{in} = \xi \kappa^2 r_{\rm in}
= 7.4_{-4.3}^{+5.9} \left(\frac{d}{10~\mathrm{kpc}}\right)\left({\cos~i}\right)^{-\frac{1}{2}} ~\mathrm{km}.
\end{equation}
where $\xi=0.41$ is a correction factor for the inner boundary
condition \citep{Kubota98}, $\kappa=1.7$ is the standard color
hardening factor \citep{Shimura95}, and the best-fit MCD model parameter 
for the periods B--D, $r_{\rm in} \sqrt{\cos i}=6.2^{+5.0}_{-3.6}$  are employed.
Since MAXI J1305$-$704 shows X-ray dip
\citep{Kennea12b, Shidatsu13}, the disk inclination angle is expected
to be $\gtsim60^\circ$ \citep{Frank87}.
%
Assuming $R_{\rm in}= 6GM/c^2$ and $i=75^\circ$, the mass
$M$ is deduced to
\begin{equation}
 M = \frac{c^{2}R_{\rm in}}{6G} = 1.7_{-1.0}^{+1.3} \left(\frac{d}{10~\mathrm{kpc}}\right)\left(\frac{\cos~i}{
\cos 75^\circ}\right)^{-\frac{1}{2}} \MO.
\label{eq1}
\end{equation}
If the accretor has a spin, the mass could be larger than the equation
above.  Figure \ref{f5} illustrates the mass-distance relation in
equation (\ref{eq1}).

Based on the past observations of X-ray outbursts from NS and BH
binaries, \citet{2003A&A...409..697M} proposed that the soft-to-hard
state transition in the outburst decay phase would presumably occur at
the 1--4\% Eddington luminosity,
$L_\mathrm{EDD}=1.25\times10^{38}~M/\MO$ $\ergs$.  The light curves
(figure~\ref{f2}) and the HID (figure~\ref{f3}) indicate that the
soft-to-hard transition occurred between period F and period G.
If the best-fit spectral model of period F continues up to the cutoff at
100~keV, the bolometric flux at the transition $F_{\rm trans}$ is
estimated to be 1.4$^{+0.1}_{-0.3}\times 10^{-9}\,\ergcms$.  In figure
\ref{f5}, the expected mass-distance relations from $4\pi d^2 F_{\rm
  trans}/L_\mathrm{EDD} =$ 1\%, and 4\% are drawn together.

Assuming $R_{\rm in}=R_{\rm ISCO}$ in the soft state periods,
$i=75^\circ$ as suggested by \citet{Shidatsu13}, and $d \sim 10$ kpc, the accretor mass is deduced 
to be $\sim 2\,\MO$ (figure \ref{f5}).
The estimated mass can be larger if the accretor has a spin.
The source distance has not been
determined. Instead, the absorption column density in the observed
X-ray spectrum against the Galactic interstellar H$_{\rm I}$ can be
used as the distance estimate.  Swift/XRT spectra obtained by multiple
observations suggest that the absorption column density is
approximately comparable to the Galactic H$_{\rm I}$ density
(eg. \citealt{Kennea12a,Miller12a,Shidatsu13}).
Thus, we suggest that MAXI J1305$-$704 would be a
non-rotating BHB located at the distance as far as the Galactic center
in the Galactic bulge, or a rotating BHB at a smaller distance
(\citealt{Shidatsu13}).

\bigskip
The authors are grateful to all members of the MAXI and the ISS-operation teams. 
This research was partially supported by the 
Ministry of Education, Culture, Sports, Science and 
Technology (MEXT), Grant-in-Aid No. 19047001, 20244015 and 24340041.


\bibliographystyle{aa}

\bibliography{ms}

\end{document}